\documentstyle[preprint,prl,aps]{revtex}

\begin{document}
\preprint{\vbox{\hbox{February 1994}\hbox{IFP-486-UNC}
\hbox{NSF-PT-94-1}\hbox{UDHEP-01-94}}}
\draft
\title{Outcome from Spontaneous \bf{$CP$} Violation for B Decays}
\author{A. W. Ackley and P. H. Frampton}
\address{Institute of Field Physics, Department of Physics and Astronomy,
University of North Carolina,\\ Chapel Hill, NC  27599-3255}
\author{B. Kayser}
\address{The Division of Physics, National Science Foundation, Washington,
D.C. 20550}
\author{C. N. Leung}
\address{Department of Physics and Astronomy, University of Delaware,
Newark, DE  19716}
\maketitle

\begin{abstract}
In the aspon model solution of the strong $CP$ problem, there is a gauged
$U(1)$ symmetry, spontaneously broken by the same vacuum expectation
value which breaks $CP$, whose massive gauge boson provides an additional
mechanism of weak $CP$ violation. We calculate the $CP$ asymmetries in
$B$ decays for the aspon model and show that they are typically smaller
than those predicted from the standard model.  A linear relation between the
$CP$ asymmetries of different decay processes is obtained.
\end{abstract}
\pacs{}

The violation of $CP$ symmetry was a surprising experimental discovery made
almost thirty years ago in the neutral kaon system \cite{CCFT}. In field
theory, one profound question about $CP$ is whether it is explicitly broken
in the fundamental lagrangian or only spontaneously broken by the vacuum.
Within the standard model {\it explicit} $CP$ violation can be accommodated
in the flavor
mixing of three families by the Kobayashi-Maskawa (KM) mechanism\cite{KM}.
The experimental information regarding $CP$ violation still comes {\it
only} from the neutral kaon system and is inadequate to determine whether
the KM mechanism is the correct underpinning of $CP$ violation.  In dedicated
$B$ studies, with more than $10^8$ samples of $B^0$ ($\bar{B^0}$) decay,
it will be possible \cite{CPasym} to test this assumption stringently
by measuring the angles of the well-known unitarity triangle whose sides
correspond to the complex terms of the equation
\begin{equation}
V_{ub}^*V_{ud} + V_{tb}^*V_{td} + V_{cb}^*V_{cd} = 0.\label{unitarity}
\end{equation}

If $CP$ is {\it spontaneously} broken, the outcome of these measurements
will be different from the standard model.  It is the purpose of the present
Letter to illustrate this in the context of the aspon model \cite{FK,FN}.

The standard model contains 19 parameters of which two, commonly denoted
by $\bar{\theta}$ and $\delta$, pertain to $CP$ violation. The value
of $\bar{\theta}$, the strong $CP$ violation parameter, is restricted by
the neutron electric dipole moment to be: $\bar{\theta} \lesssim 2\times
10^{-10}$.  The KM mechanism offers no solution of this fine-tuning which
is generally explained by an independent mechanism.  In the aspon model
which solves the strong $CP$ problem, there is a new gauged $U(1)$
symmetry which is spontaneously broken by the same vacuum expectation
value that breaks $CP$.  The resulting massive gauge boson, the aspon,
provides an additional mechanism for weak $CP$ violation.

The three angles of the unitarity triangle (conventionally defined as
$\alpha$, $\beta$, $\gamma$ between the first and second, second and third,
and third and first sides in (\ref{unitarity}), respectively) can be
separately measured for the standard model by the time-dependent $CP$
asymmetry \cite{Nir},
\begin{equation}
a_f(t) = {{\Gamma (B^0(t) \rightarrow f) - \Gamma (\bar{B^0}(t) \rightarrow f)}
\over {\Gamma (B^0(t) \rightarrow f) + \Gamma (\bar{B^0}(t) \rightarrow f)}}
\end{equation}
where the final state $f$ is a $CP$ eigenstate. We define $q$, $p$ in
$B^0$--$\bar{B^0}$ mixing by the mass eigenstates $B_{1,2}$:
\begin{equation}
| B_{1,2} \rangle = p | B^0 \rangle \pm q | \bar{B^0} \rangle
\end{equation}
and similarly for $K_{1,2}$ in the kaon system.  $A$, $\bar{A}$ are the
decay amplitudes:
\begin{equation}
A,\bar{A} = \langle f | H | B^0, \bar{B^0} \rangle
\end{equation}
Let us consider the specific cases of  $f = {\pi}^+{\pi}^-, {\psi}{K_S}$
from $B_d$ decay and  $f = {\rho}{K_S}$ from $B_s$ decay. We define
$\lambda(f)$ by
\begin{eqnarray}
\lambda({\pi}^+{\pi}^-) & = & ~\left({q \over p}\right)_{B_d} \left(
{\bar{A} \over A}\right)_{B_d\rightarrow{\pi}^+{\pi}^-} \label{pipi} \\
\lambda(\psi K_S)~ & = & ~\left({q \over p}\right)_{B_d} \left({\bar{A} \over
A}\right)_{B_d\rightarrow \psi K} \left({q \over p}\right)_K \label{psik} \\
\lambda(\rho K_S)~ & = & ~\left({q \over p}\right)_{B_s}\left({\bar{A} \over
A}\right)_{B_s\rightarrow{\rho}K}\left({q \over p}\right)_K^*\label{rhok}
\end{eqnarray}
The complex conjugate appears in (\ref{rhok}) because $B^0_s \rightarrow
\bar{K^0}$ while $B^0_d \rightarrow K^0$.  If to a sufficiently good
approximation $|q/p| = 1$, and $|\bar{A}/A| = 1$ as we shall show for
the aspon model below, then $\lambda(f)$ is related to the $CP$ asymmetry
through the $B_1-B_2$ mass difference $\Delta M$ by
\begin{equation}
a_f(t) = - \text{Im } \lambda(f) \sin(\Delta Mt). \label{expt}
\end{equation}
In the standard model the angles of the unitarity triangle are related to the
$\lambda(f)$ by \cite{Nir}:  $\sin 2\alpha = \text{Im }\lambda
({\pi}^+{\pi}^-)$, $\sin 2\beta = -\text{Im }\lambda(\psi K_S)$, and
$\sin 2\gamma = -\text{Im }\lambda(\rho K_S)$.  Such relations are no longer
valid in the aspon model because Im$(q/p)_{B_d}$ has a major contribution
from aspon exchange and Im$(q/p)_K$ is dominated by aspon exchange.

To evaluate the $CP$ asymmetries in $B$ decays for the aspon model we
need to evaluate the different factors in the $\lambda(f)$ given in
Eqs.\ (\ref{pipi}--\ref{rhok}) above. More precisely we need, from
Eq.\ (\ref{expt}), the imaginary part of the $\lambda(f)$. The aspon model
adds new Feynman diagrams involving aspon exchange to those involving
$W$ exchange already present in the standard model. Because $CP$ is only
spontaneously broken, the $W$-exchange amplitudes are predominantly real
and have very small phases while the aspon-exchange has a much smaller
magnitude but an unpredicted arbitrary phase. As a result, the $|\text{Im
}\lambda
(f)|$ appearing in Eq.\ (\ref{expt}) are of order $0.002$ or less
compared to the standard model expectation that $|\text{Im }\lambda(f)|$ be
generally of order of, although less than, unity. Thus, $CP$ asymmetries
in $B$ decays are predicted to be correspondingly smaller than in the
standard model. This is our principal result. The remainder of this
Letter provides more technical details.

Firstly, we examine the three mixing factors $(q/p)$ for the $B_d$, $B_s$,
and $K$ neutral meson systems.  The definition is
\begin{equation}
\left({q\over p}\right)_{\xi} = \left({M_{21}(\xi)\over M_{12}(\xi)}
\right)^{\frac12}
\end{equation}
where $M_{12}(\xi)$ and $M_{21}(\xi)$ are the amplitudes for
$\bar{\xi} \rightarrow \xi$ and $\xi \rightarrow \bar{\xi}$, respectively.
The quantity $M_{12}(\xi)$
is the sum of one-loop $W^+W^-$ exchange and tree-level aspon exchange
amplitudes.  For example,
\begin{equation}
M_{12}(B_d) \propto \frac{G_F m_t^2}{32 \pi^2} f(z_t) \left(V_{td}^*V_{tb}
\right)^2 \frac{g^2}{M_W^2} + \left(x_3^*x_1\right)^2 \frac{g_A^2}{M_A^2}
\label{bd}
\end{equation}
Similarly,
\begin{eqnarray}
M_{12}(B_s) & \propto & \frac{G_F m_t^2}{32 \pi^2} f(z_t) \left(
V_{ts}^*V_{tb} \right)^2 \frac{g^2}{M_W^2} + \left(x_3^*x_2\right)^2
\frac{g_A^2}{M_A^2} \\
M_{12}(K) & \propto & \frac{G_F m_c^2}{32 \pi^2} \left(V_{cd}^*V_{cs}\right)^2
\frac{g^2}{M_W^2} + \left(x_2^*x_1\right)^2 \frac{g_A^2}{M_A^2} \label{k}
\end{eqnarray}
where $m_t$ ($m_c$) is the mass of the top (charm) quark, $z_t=m_t^2/M_W^2$ and
\begin{equation}
f(z) = \frac{1}{4} \left[ 1+\frac9{1-z}-\frac6{(1-z)^2}-\frac{6z^2\ln
z}{(1-z)^3} \right] \label{box}
\end{equation}
The function $f(z_t)$ takes values between $0.64$ and $0.51$ for $m_t$
between $120$ and $200$ GeV.  In Eqs.\ (\ref{bd}--\ref{k}), $g$ is the
coupling constant for electroweak $SU(2)$, $g_A$ is the coupling constant
for the new gauged $U(1)$, $M_A$ is the mass of the aspon, and the $x_i$
are given in the notation of \cite{FN} as $x_i = F_i / M$.  The
expression for $K^0$--$\bar{K^0}$ mixing uses the fact that
$|V_{cd}^*V_{cs}m_c| \gg |V_{td}^*V_{ts}m_t|$ and hence amongst possible
internal quarks the charm quark dominates the top quark, unlike the case
for $B^0$--$\bar{B^0}$ mixing where the top quark dominates.  This is
important because, as a consequence, aspon exchange dominates the
imaginary part of $M_{12}(K)$.  This allowed upper limits on the aspon mass
(and the $CP$ violating scale) to be arrived at in the original aspon
papers\cite{FK,FN}.  The aspon mass was made sufficiently small so that
the $K^0$--$\bar{K^0}$ mixing phase compensated the smallness of the
phase in the decay amplitude for $s \rightarrow u\bar{u}d$ ($W$ exchange).
The difference in physics between the neutral $B$ and
$K$ systems depends further on the inequalities $|x_3|\ll |x_2|\ll
|x_1|$.

In the neutral $B$ system the phase of the amplitude for the decay $b
\rightarrow u\bar{u}d$ ($W$ exchange) is again smaller in the aspon model
than in the standard model in a suitable phase convention.  But in this
case the phase of the $B^0$--$\bar{B^0}$ mixing has significant
contributions from $W^+W^-$ exchange. As can be gleaned from
Eqs.\ (\ref{bd}--\ref{k}), there are two numerical reasons, both acting
in the same direction, why the imaginary part of the one-loop
$W^+W^-$ exchange is comparable to that of the tree-level aspon exchange in
$B^0$--$\bar{B^0}$ mixing.  Firstly, the $W^+W^-$ exchange amplitude in
$B^0$--$\bar{B^0}$ mixing is enhanced by a relative factor $(m_t/m_c)^2$;
secondly, there is a suppression $|x_3/x_2|^2$ (for $B_d$) in the aspon
exchange amplitude relative to the neutral kaon system.

To evaluate the mixing we need the aspon model values of the $V_{ij}$.
These are given by \cite{FN}:
\begin{equation}
V_{ij} = \left(j_L^{\dag}\right)_{il} C_{ln} \left(k_L\right)_{nj} +
\tilde{x}_i x_j^*
\end{equation}
where $C_{ln}$ is a real matrix.  Up to quadratic order in the $x_i$,
$\tilde{x}_i$, and using $m_d \ll m_s \ll m_b$, $m_u \ll m_c \ll m_t$,
the matrices $j_L^{\dag}$ and $k_L$ are given by
\begin{equation}
j_L^{\dag} \simeq \left( \begin{array}{ccc}
 \tilde{a}_1               & 0                         & 0 \\
 -\tilde{x}_1^*\tilde{x}_2 & \tilde{a}_2               & 0 \\
 -\tilde{x}_1^*\tilde{x}_3 & -\tilde{x}_2^*\tilde{x}_3 & \tilde{a}_3
\end{array} \right)
\end{equation}
and
\begin{equation}
k_L \simeq \left( \begin{array}{ccc}
 a_1 & -x_1x_2^* & -x_1x_3^* \\
 0   & a_2       & -x_2x_3^* \\
 0   & 0         & a_3
\end{array} \right)
\end{equation}
In these expressions $a_i = 1 - \frac12 |x_i|^2$, $\tilde{a}_i =
1 - \frac12 |\tilde{x}_i|^2$, and $\tilde{x}_i = C_{ij}x_j$, where
the overtilde refers to variables in the charge $(+\frac23)$ sector
as opposed to the charge $(-\frac13)$ sector.

The $\bar{B^0}$ decay amplitudes are given by
\begin{eqnarray}
\bar{A}_{B_d\rightarrow{\pi}^+{\pi}^-} & \propto & 3\left( V_{ud}^*V_{ub}
\right) \frac{g^2}{M_W^2} + x_3^*x_1|\tilde{x}_1|^2\frac{g_A^2}{M_A^2}
\zeta_{\pi \pi} \label{bpp} \\
\bar{A}_{B_d\rightarrow\psi K} & \propto & \left( V_{cs}^*V_{cb}\right)
\frac{g^2}{M_W^2} + 3 x_3^*x_2|\tilde{x}_2|^2\frac{g_A^2}{M_A^2}
\zeta_{\psi K} \label{bpk} \\
\bar{A}_{B_s\rightarrow\rho K} & \propto & \left( V_{ud}^*V_{ub}\right)
\frac{g^2}{M_W^2} + x_3^*x_1|\tilde{x}_1|^2\frac{g_A^2}{M_A^2}\zeta_{\rho K}
\label{brk}
\end{eqnarray}
where the first and second terms originate from $W$ exchange and aspon
exchange, respectively.  The $\zeta$ factors, which are of order 1,
account for the fact that the aspon exchange amplitudes involve
different strong interaction dynamics from the $W$-exchange amplitudes.
The factors of $3$ in Eqs.\ (\ref{bpp}--\ref{bpk})
arise from color $SU(3)$; the absence of a relative color weighting in
Eq.\ (\ref{brk}) follows from considerations of isospin.
In all cases, the real and imaginary parts are
dominated by the $W$-exchange amplitudes and so $|\bar{A}/A| = 1$.  This
confirms the validity of Eq.\ (\ref{expt}).

We may parameterize $x_j = |x_j| \exp(i{\phi}_j)$.  The phases ${\phi}_j$
are unknown.  If we focus on the imaginary parts of the quantities in
Eqs.\ (\ref{pipi}--\ref{rhok}) and assume that $g_A = g$, we find
\begin{eqnarray}
\text{Im}\left(\frac{q}{p}\right)_K & = & \frac{2}{V_{cd} V_{cs}}
\lbrace[(V_{ud} V_{cs} - V_{us} V_{cd})^2 - V_{cs}^2] |x_1 x_2|
\sin (\phi_1 - \phi_2) \nonumber \\
& & \mbox{} + V_{us} V_{ub} |x_1 x_3| \sin (\phi_1 - \phi_3) \nonumber \\
& & \mbox{} - V_{ud} V_{ub} |x_2 x_3| \sin (\phi_2 - \phi_3)\rbrace
\nonumber \\
& & \mbox{} - \frac{32 \pi^2}{G_F m_{c}^2} \frac{M_{W}^2}{M_{A}^2}
{|x_1 x_2|^2 \over (V_{cd} V_{cs})^2} \sin 2(\phi_1 - \phi_2) \label{imk} \\
\text{Im}\left(\frac{q}{p}\right)_{B_d} & = & 2 \frac{V_{td}}{V_{tb}}
|x_1 x_3| \sin (\phi_1 - \phi_3) \nonumber \\
& & \mbox{} + 2 \frac{V_{ts}}{V_{tb}} |x_2 x_3| \sin (\phi_2 - \phi_3)
\nonumber \\
& & \mbox{} - \frac{32 \pi^2}{G_F m_{t}^2 f(z_t)} \frac{M_{W}^2}{M_{A}^2}
{|x_1 x_3|^2 \over (V_{td} V_{tb})^2} \sin 2(\phi_1 - \phi_3) \label{imbd} \\
\text{Im}\left(\frac{q}{p}\right)_{B_s} & = & - 2 \frac{V_{td}}{V_{ts}}
|x_1 x_2| \sin (\phi_1 - \phi_2) \nonumber \\
& & \mbox{} + 2 \frac{V_{td}}{V_{tb}} |x_1 x_3|
\sin (\phi_1 - \phi_3) \nonumber \\
& & \mbox{} + 2 \frac{V_{ts}}{V_{tb}} |x_2 x_3|
\sin (\phi_2 - \phi_3) \nonumber \\
& & \mbox{} - \frac{32 \pi^2}{G_F m_{t}^2 f(z_t)} \frac{M_{W}^2}{M_{A}^2}
{|x_2 x_3|^2 \over (V_{ts} V_{tb})^2} \sin 2(\phi_2 - \phi_3) \label{imbs} \\
\text{Im}\left(\frac{\bar{A}}{A}\right)_{B_d\rightarrow\pi^+\pi^-} & = &
\text{Im}\left(\frac{\bar{A}}{A}\right)_{B_s\rightarrow\rho K}
\nonumber \\
& = & 2 \frac{V_{us}}{V_{ud}} |x_1 x_2| \sin (\phi_1 - \phi_2)
+ 2 \frac{V_{ub}}{V_{ud}} |x_1 x_3| \sin (\phi_1 - \phi_3) \label{uud} \\
\text{Im}\left(\frac{\bar{A}}{A}\right)_{{B}_d\rightarrow\psi K} & = &
\frac{2}{V_{cs} V_{cb}} \lbrace - V_{td} V_{tb} |x_1 x_2|
\sin (\phi_1 - \phi_2) \nonumber \\
& & \mbox{} + V_{td} V_{ts} |x_1 x_3| \sin (\phi_1 - \phi_3) \nonumber \\
& & \mbox{} + [(V_{cs} V_{tb} - V_{cb} V_{ts})^2 - V_{cs}^2] |x_2 x_3|
\sin (\phi_2 - \phi_3)\rbrace \label{ccs}
\end{eqnarray}
In Eqs.\ (\ref{imk}--\ref{imbs}), the terms of order $|x_i x_j|$ are from the
one-loop $W^+W^-$ exchange and the terms of order $|x_i x_j|^2$ are from
tree-level aspon exchange.  The order $|x_i x_j|^2$ contributions from the box
diagrams have been ignored because in general these are much smaller than the
aspon contributions displayed.  In Eqs.\ (\ref{uud}--\ref{ccs}), the
displayed terms are from $W$ exchange only because, as mentioned earlier,
contributions from aspon exchange are negligible.  To the order of
accuracy shown in Eqs.\ (\ref{imk}--\ref{ccs}), the $V_{ij}$ can be
regarded as real and are given by the corresponding $C_{ij}$.

Combining Eqs.\ (\ref{imk}--\ref{ccs}) with Eqs.\ (\ref{pipi}--\ref{rhok})
yields the results
\begin{mathletters} \label{mag}
\begin{eqnarray}
|\text{Im } \lambda({\pi}^+{\pi}^-)| & \lesssim & 1 \times 10^{-5}\\
|\text{Im } \lambda(\psi K_S)| & \lesssim & 2 \times 10^{-3}\\
|\text{Im } \lambda(\rho K_S)| & \lesssim & 2 \times 10^{-3}
\end{eqnarray}
\end{mathletters}
The numerical values were obtained by using the central values of the
$V_{ij}$ listed in the most recent Review of Particle Properties\cite{PDG}
and the value $300$ GeV for the aspon mass, and the magnitudes of the
$x_i$ have been restricted by the upper limit on the strong $CP$ parameter
$\bar{\theta}$ to be \cite{FN}: $|x_1|\lesssim 10^{-2}$,
$|x_2|\lesssim 10^{-3}$, $|x_3|\lesssim 10^{-4}$.  The resulting $CP$
asymmetries $a_f(t)$ are in general smaller than those predicted by the
standard model.  We note that the final states with a kaon typically have
larger asymmetries. If $CP$ violation is not detected in the planned $B$
studies as initially envisaged, it may be a signal that spontaneous $CP$
violation is at work.

It is also interesting to observe from Eqs.\ (\ref{pipi}--\ref{rhok})
that, since the aspon exchange terms in Eqs.\ (\ref{bpp}--\ref{brk}) are
negligible,
\begin{eqnarray}
\frac{\lambda(\psi K_S) \lambda(\rho K_S)}{\lambda({\pi}^+{\pi}^-)}
& = & \left(\frac{q}{p}\right)_{B_s}
\left(\frac{\bar{A}}{A}\right)_{B_d\rightarrow\psi K}
\nonumber \\
& = & \left(\frac{q}{p}\right)_{B_s}
\left(\frac{\bar{A}}{A}\right)_{B_s\rightarrow{D_s}^+{D_s}^-}
\nonumber \\
& = & \lambda({D_s}^+{D_s}^-) .
\label{smrel}
\end{eqnarray}
In the aspon model where the $\lambda(f)$ have unit moduli and
$|\text{Im } \lambda(f)| \ll 1$, this relation implies a
linear relation for the imaginary parts:
\begin{equation}
\text{Im } \lambda(\psi K_S) + \text{Im } \lambda(\rho K_S) -
\text{Im } \lambda({\pi}^+{\pi}^-) - \text{Im } \lambda({D_s}^+{D_s}^-)
= 0 , \label{asponrel}
\end{equation}
which provides an additional test of the aspon model.  We may also infer
from Eqs.\ (\ref{mag}) and (\ref{asponrel}) that
\begin{equation}
|\text{Im } \lambda({D_s}^+{D_s}^-)| \lesssim 4 \times 10^{-3}.
\end{equation}

In conclusion, our result is that if the aspon model were correct, $CP$
asymmetries in $B$ decays would be much smaller than predicted by the
standard model and the relation (\ref{asponrel}) would be satisfied.
Although we have considered only
final states which are $CP$ eigenstates, it is expected that the $CP$
violation effects will likewise be small for the $CP$ non-eigenstates.

If $CP$ violation in the neutral $B$ meson decays shows up at the level
expected from the KM mechanism, it will disfavor spontaneous $CP$ violation.
On the other hand, if this is {\it not} observed, spontaneous $CP$ violation,
as exemplified by the aspon model, will become a viable alternative to the
KM mechanism.  To verify then the aspon model would require the ability
to measure very small $CP$ asymmetries in the $B$ system as well as
detection of the aspon in a hadron collider \cite{FKNW}.

This work was supported in part by the U.S. Department of Energy under
Grants DE-FG05-85ER-40219 and DE-FG02-84ER-40163.

\end{document}